# Mechanism of thermal renaturation and hybridization of nucleic acids:

# Kramers' process and universality in Watson-Crick base pairing

*Jean-Louis Sikorav*[†], *Henri Orland*[†], *and Alan Braslau*[‡]

DSM, Institut de Physique Théorique, IPhT, CNRS, MPPU, URA2306; DSM, Service de Physique de l'État Condensé, CNRS URA2464, CEA/Saclay, F-91191 Gif-sur-Yvette, France

Email: jean-louis.sikorav@cea.fr

*To whom correspondence should be addressed. Email: jean-louis.sikorav@cea.fr

[†]Institut de Physique Théorique

[‡]Service de Physique de l'État Condensé

ABSTRACT Renaturation and hybridization reactions lead to the pairing of complementary single-stranded nucleic acids. We present here a theoretical investigation of the mechanism of these reactions *in vitro* under thermal conditions (dilute solutions of single-stranded chains, in the presence of molar concentrations of monovalent salts and at elevated temperatures). The mechanism follows a Kramers' process, whereby the complementary chains overcome a potential barrier through Brownian motion. The barrier originates from a single rate-limiting nucleation event in which the first complementary base pairs are formed. The reaction then proceeds through a fast growth of the double helix. For the DNA of bacteriophages T7, T4 and $\varphi$X174 as well as for *Escherichia coli* DNA, the bimolecular rate $k_2$ of the reaction increases as a power law of the average degree of polymerization $<N>$ of the reacting single-strands: $k_2 \propto <N>^\alpha$. This relationship holds for $100 \leq <N> \leq 50,000$ with an experimentally determined exponent $\alpha = 0.51 \pm 0.01$. The length dependence results from a thermodynamic excluded-volume effect. The reacting single-stranded chains are predicted to be in universal good solvent conditions, and the scaling law is determined by the relevant equilibrium monomer contact probability. The value theoretically predicted for the exponent is $\alpha = 1-\nu\theta_2$, where $\nu$ is Flory's swelling exponent ($\nu \approx 0.588$) and $\theta_2$ is a critical exponent introduced by des Cloizeaux ($\theta_2 \approx 0.82$), yielding $\alpha = 0.52 \pm 0.01$, in agreement with the experimental results.

**Introduction**

Genetic information is embedded in polymerized nucleic acids and, as a result, many topics of genetics raise issues of polymer physics. The broad field of hybridization and renaturation of nucleic acids is one instance of this situation. These magnificent sequence-specific reactions rely on the process of complementary recognition through Watson-Crick base pairing (figure 1). They are involved in essential biological processes such as genetic recombination and RNA interference, amongst others. Beyond their biological importance, these reactions constitute *the* basic tool of biotechnologies, whether in PCR (Polymerase Chain Reaction), DNA chips, molecular beacons or DNA computing. Understanding their mechanism is therefore of considerable interest from both fundamental and applied viewpoints.

Hybridization and renaturation of nucleic acids have been investigated in different experimental systems, ranging from homogeneous to heterogeneous, both in the absence of proteins as well as in the presence of proteic catalysts (see references[1-3] for reviews). The most thorough studies have been performed in the bulk of homogeneous aqueous solutions under fairly large concentrations of monovalent salts and at high temperatures. Such standard conditions are known as thermal renaturation or hybridization conditions[4]. They are optimal in terms of yields[5,6] and lead to the complete annealing of even very long complementary strands (tens of kilobases long[7,8]). The specificity of the reaction is exquisite, allowing the detection of a single complementary sequence among more than $10^{10}$ non homologous chains[9].

How thermal renaturation and hybridization reactions of long chains proceed is still poorly understood. Two different mechanisms have been considered: the first based on a diffusion-controlled encounter process and the second on a thermodynamic excluded-volume effect. Both are believed to account for the experimental results, in particular for the length dependence of DNA renaturation, as described forty years ago by Wetmur and Davidson[7]. These authors showed that the rate of the reaction, defined by a bimolecular rate constant $k_2$, increases as a power law of the average degree of polymerization of the complementary single-strands

$$k_2 \propto <N>^\alpha \qquad (1)$$

with $\alpha \approx 0.5$. This law is observed to hold for $100 \leq <N> \leq 50,000$, over more than two orders of magnitude in $<N>$. Wetmur and Davidson[7] claimed that this dependence on the square root of the length is compatible with both diffusion control and excluded volume mechanisms. For both cases, their reasoning relies on the crucial *ad hoc* assumption that the reacting single-strands have a Gaussian conformation, thus being precisely at Flory's $\Theta$ point[10,11]. They then rejected the possibility of a diffusion control on the grounds that the measured rates were quantitatively much smaller than those predicted by the Smoluchowski equation for spherical molecules[12]. However, diffusion-controlled reactions can have much smaller rates due to geometrical constraints[13]. This led Schmitz and Schurr[14] to challenge their rejection and to assert that thermal DNA renaturation was indeed diffusion-controlled. This forty-year-old problem has remained unresolved. In the present work, we try to reach a better understanding of the mechanism of thermal renaturation and hybridization reactions by taking into account the current state of knowledge of polymer physics and reaction theory.

The presence or absence of an activation barrier provides a basic criterion for the classification of chemical reactions. A reaction that proceeds unimpeded by an energy barrier is called diffusion-controlled. The theory of diffusion-controlled encounter reactions is based on an understanding of Brownian fluctuations and was pioneered by Smoluchowski[12]. Diffusion-controlled encounter reactions constitute an important topic of physical chemistry, especially in the life sciences[15-18].

The theory of reactions requiring the crossing of one (or more) energy barrier(s) was developed initially by van't Hoff and Arrhenius. The state found at the top of the highest barrier is called the activated complex or transition state. Thus, the Transition State Theory (TST), which assumes the existence of a thermodynamic equilibrium between the reactants and this transition state[19], has been for a long time the dominant theory used to analyze the crossing of an energy barrier. It was in this context that Wetmur and Davidson[7] analyzed the mechanism of DNA renaturation. An alternative approach to the problem of barrier crossing involving Brownian fluctuations was developed in 1940 by Kramers[20]. The recognition of the importance of the Kramers' approach for reaction rate theory has been steadily growing[21], it is of particular importance for reactions involving polymers[22] or biopolymers[23-26], and its relevance for the understanding of renaturation and hybridization reactions will be illustrated here. Using modern concepts of polymer dynamics, we will show that the published experimental data are, in fact, incompatible with both TST and a diffusion-controlled encounter mechanism and that the reaction can only be described as a Kramers' process.

*Background knowledge on thermal renaturation and hybridization reactions*

Renaturation and hybridization are complex reactions described in detail both in reviews[1,2,27,28] and in textbooks[29,30]. Hybridization reactions can be performed with non-stoichiometric amounts of complementary single-strands, in particular, with a vast excess of one of the two strands. Under the situation where the complementary sequences are present in equal amounts, the mechanisms of hybridization and of renaturation of long complementary chains (made of 100 monomers or more) are essentially identical and can be discussed as for the case of the simpler renaturation reaction.

A general procedure to investigate DNA renaturation begins with linear, double-stranded DNA chains extracted from a given cell or virus. These chains, denoted *H* (for helical), are denatured at high temperature (well above the melting temperature $T_m$) to generate complementary single strands $S$ and $\bar{S}$ in equal amounts: $H \rightarrow S + \bar{S}$.

Following this denaturation, the sample is usually thermally quenched. Thermal renaturation of DNA takes place in a pH neutral, aqueous solution, in the presence of rather large concentrations of monovalent salts (typically about 1 M NaCl) and at elevated temperature (typically around 65-70°C, thus 20-25°C below the melting temperature $T_m$). The renaturation reaction is initiated by the addition of the salt, thus decreasing the electrostatic repulsion between the like-charged chains. It is monitored by an appropriate technique (see reference[2] for review) such as UV absorbance at 260 nm[4,6,7,31], sedimentation[8], hydroxyapatite binding[32,33] or fluorescence spectroscopy[34].

Let $k_2(t)$ denote the bimolecular rate for the reverse reaction $S + \bar{S} \rightarrow H$. Under the experimental conditions of thermal renaturation (20-25°C below $T_m$), it is irreversible and the rate equation is given by

$$\frac{d[H]}{dt} = k_2[S][\bar{S}] = k_2[S]^2 \qquad (2)$$

since $[S] = [\bar{S}]$. The conservation of matter yields $[H] + [S] = C_0$, where $C_0$ is the initial concentration in helical chains. The bimolecular rate $k_2(t)$ is expected on theoretical grounds to have a complex temporal evolution, in particular with respect to the initial and asymptotic regimes. Fortunately, a simpler form of equation 2 is found to describe the rate of the reaction in a broad intermediate time scale:

*1) Order of the reaction*

Assuming that $k_2$ is constant over time $t$, one obtains

$$\frac{1}{f_{ss}} = k_2 C_0 t + 1 \qquad (3)$$

where $f_{SS} = \frac{[S]}{C_0}$ is the fraction of DNA remaining single-stranded at time *t*. Equation 3, which predicts that the reciprocal of the fraction of DNA remaining single-stranded increases linearly with time, describes the temporal evolution of the reaction in an intermediate regime, between 10 to 10,000 seconds and up to 75-90% renaturation[6,7]. This intermediate regime is followed by a slowing down of the reaction which is due to the maintained reactivity of the dangling single-stranded tails formed at the beginning of the reaction[35-37]. (These reactions have been studied using complementary single-strands of varying chain lengths obtained by a random degradation process and thus having partially overlapping sequences. The first encounter of two complementary strands produces duplex species that are only partially double-stranded, and the single-stranded tails in these species remain accessible for further annealing, but with a decreased reactivity which leads to the observed slowing down of the reaction.) These late events do not affect the intermediate, linear regime. This bimolecular reaction is therefore kinetically second-order with respect to time in this regime.

*2) Dependence of the rate on the chain concentration*

Equation 3 is also valid with respect to changes in the initial chain concentration $C_0$, provided that the system remains dilute[6,7,31]. Increasing the chain concentration beyond a certain limit leads to a different regime that does not follow second-order kinetics: $k_2$ is not a constant but is observed to increase by a factor of two for a tenfold increase in DNA concentration[8]. This, in fact, corresponds to entering the semi-dilute regime where the single-stranded chains begin to overlap[38] and will not be discussed further here. We will consider the case where the reaction takes place in a dilute solution of single-stranded chains and is kinetically second-order, both with respect to time and to chain concentration.

*3) The nucleation and zipping mechanism*

The reaction is thought to proceed in two steps. A nucleation event first leads to the formation a few correct base pairs between the two complementary strands and is followed by a zipping, irreversible reaction during which the rest of the base pairing proceeds. The fact that the reaction follows second-order kinetics implies that the zipping step, which is monomolecular, is very fast. The expected rapidity of the second step agrees with the data obtained by relaxation spectroscopy for zipping, showing growth rates in the order of $10^6$-$10^7$ base pairs per second[39,40]. The bimolecular nucleation event is therefore the rate-limiting step.

The events preceding the nucleation are not precisely understood. However it is known from experiments that the formation of specific, stable hybrids between ribo- or deoxyriboligonucleotides and long complementary DNA in thermal conditions requires oligonucleotides of about 15 bases or more[41,42]. The same length requirement is also expected to be present in renaturation and hybridization of two long complementary single-strands. According to Manning[43] there must exist a nucleus consisting of separated (unbonded) and aligned complementary bases of about this length (13-16 bases as explained below), and the nucleation event occurs with the formation of one or several (consecutive) base pairs within this nucleus. A major feature of this mechanism is the assumption of the reversibility of the reaction between the separated strands and the nucleus with the existence of a thermodynamic equilibrium between the two states:

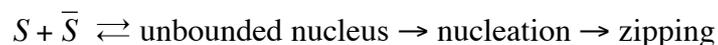

$$S + \bar{S} \rightleftarrows \text{unbounded nucleus} \rightarrow \text{nucleation} \rightarrow \text{zipping}$$

The existence of this equilibrium manifests itself through a law of mass action behavior. We shall see that both salt concentration and length dependences of the reaction rate are indications of such a behavior.

*The relation between the nucleation rate constant and the overall bimolecular rate constant for unique sequences*

The overall bimolecular rate constant $k_2$ is expected to be proportional to the product of a nucleation rate constant $k_{nu}$ times the number of possible nucleation sites per chain. The number of nucleation sites is in turn a function of the kinetic complexity of the DNA chains employed. The kinetic complexity $N_{kin}$ of an $N$ base-pair long DNA molecule is given by the number of base pairs present in non-repeating sequences in a renaturation/hybrization experiment. A DNA chain devoid of repeated sequences in a given experiment is called unique or simple[7,28], and for a unique sequence of length $N$, $N_{kin} = N$. Note that the kinetic complexity is not an intrinsic property of the molecule considered but is expected to depend on the annealing conditions. The viral and bacterial DNA sequences considered here are, in fact, known to be unique under thermal conditions. For such unique sequences, one expects the number of nucleation sites to be proportional to the average degree of polymerization of the denatured strands $<N>$. Thus,

$$k_2 \propto k_{nu}<N> \qquad (4)$$

The scaling law for $k_2$ (as $<N>^\alpha$, equation 1) predicts that the net nucleation rate constant $k_{nu}$ decreases with increasing $<N>$:

$$k_{nu} \propto <N>^{\alpha-1} \qquad (5)$$

This scaling can be understood from the polymeric nature of single-stranded DNA, and this will be developed below.

*4) Effects of temperature, ionic strength, and viscosity*

a) The *apparent* renaturation rate constant measured at a constant ionic strength shows a broad peak (or "bell-shaped" dependence on temperature), with a maximum between 15 to 30° C below the melting temperature[4,7,8]. A similar dependence was first observed for the hybridization of complementary homopolynucleotides[44,45]. The qualitative and quantitative understanding of these curves requires to take into account both renaturation and denaturation reactions, occurring simultaneously. Far below the melting temperature (in particular 20-25°C below $T_m$), the equilibrium between the double-stranded helix and the separated single-strands is largely shifted in favor of the double helix, and the denaturation process can be neglected. When the temperature increases, however, the extent of renaturation decreases, from close to 100% at 20-25°C below $T_m$ to 50% (by definition) at $T_m$. Equation 2 ceases to be valid and must be replaced by:

$$\frac{d[H]}{dt} = k_2[S]^2 - k_{-1}[H] \qquad (6)$$

where $k_{-1}$ is the denaturation rate constant. Taking this correction into account, Thrower and Peacocke[5,6] found that the bimolecular rate constant $k_2$ actually increases with temperature up to $T_m$ (the statement often encountered in the literature that $k_2$ is equal to zero at $T_m$ is erroneous, as is the affirmation that thermal renaturation optimizes the rate of the reaction; rather the yield is optimized at about 20-25°C below $T_m$). Thrower and Peacocke[5,6] furthermore determined the activation energy of DNA renaturation from an Arrhenius equation. The energy of activation $\Delta G^*$ obtained from experiments carried out at temperatures between 60 and 90°C was found to be identical, $\Delta G^* = 30 \pm 4$ kJ.mol$^{-1}$, for three NaCl concentrations (0.4, 0.6 and 1 M).

b) The effects of the ionic strength on the renaturation reaction arise from the polyelectrolytic character of the reagents. The two negatively-charged polymers strongly repel each other at low salt concentrations, and adding salt decreases this repulsion. At a constant temperature, $k_2$ is found to increase as a power law of the activity $a_{Na}$ of the sodium ions, studied between 20 and 200 mM NaCl[8]:

$$k_2 \propto a_{Na}^x \qquad (7)$$

with an exponent $x$ varying from $x = 3.7$ at 25°C to $x = 2.5$ at 65°C. At the molecular level, the pairing of the complementary single-strands is associated with the concomitant uptake of sodium ions, and this power law reflects the law of mass action, as expected for a thermodynamic equilibrium involving polyelectrolytes and salts[43,46]. The data obtained by Studier[8] were analyzed in the framework of the theory of Manning-Oosawa condensation by Manning[43] who concludes, quantitatively, that 13-16 bases are involved in the unbonded nucleus preceding the nucleation event. This value agrees with the estimate given for the stability of oligonucleotide hybridization, as stated above. At higher concentrations the rate increase levels off and the power law breaks down. No plateau is observed (up to 3.2 M NaCl in the data of Wetmur and Davidson[7]) and these results at high monovalent salt concentrations are inconclusive regarding the mechanism of the reaction under thermal conditions.

c) Lastly, the rate constant is found to be inversely proportional to the viscosity $\eta$ of the solution[6,7,47]. This dependence was determined for a variety of viscogenic compounds, taking into account as well their effect on the stability of the double helix[7] and holds over a broad range of temperature between $T_m - 30°C$ and $T_m - 5°C$ [47]. The inverse proportionality of the rate constant to the viscosity is a crucial observation as it points to a role of the solvent fluctuations in the reaction.

*5) The scaling law $k_2 \propto <N>^{\alpha}$*

The length dependence of the reaction has been the subject of several investigations carried out with randomly sheared nucleic acids. A compilation of published data for renaturation and hybridization rates as a function of single-stranded nucleic sizes is shown in figure 2a. Five sets of data are plotted: those of Wetmur and Davidson[7] for the renaturation of bacteriophage T7 (up triangles) and T4 (down triangles) DNA at 1 M Na$^+$, of Hinnebusch *et al*.[33] for the renaturation of *Escherichia coli* DNA (diamonds) at 0.18 M Na$^+$, and of Hutton and Wetmur[48] for bacteriophage $\varphi$X174 DNA-RNA hybrids (squares) as well as the renaturation of bacteriophage T7 DNA (circles) at 0.4 M Na$^+$. The data are drawn here with the rate constant $k_2$ expressed in units of $\text{mole}_{\text{chain}}^{-1}\text{sec}^{-1}$ (rather than $\text{mole}_{\text{phosphate}}^{-1}\text{sec}^{-1}$) versus <$N$>. A power law $k_2 = A_i \langle N \rangle^\alpha$ with a common exponent $\alpha = 0.502 \pm 0.009$ can be used to describe the different measurements (solid lines in figure 2a), in excellent agreement with the previous results published separately.

The data gathered in figure 2a correspond to different experimental conditions (salt concentration and temperature) and techniques (UV absorbance and hydroxyapatite chromatography), as well as nucleic acids of different origins and types (DNA, glucosylated for T4, and RNA). These data are rescaled in figure 2b correcting for these differences:

1) The salt concentration dependence (Wetmur and Davidson[7], table 6 and Studier[8], figure 3) leads to multiplying the data at 0.4 M Na$^+$ (circles) by a factor $k_2^{1M}/k_2^{0.4M} = 123/69.3$ and the data at 0.18 M Na$^+$ (diamonds) by a factor $k_2^{1M}/k_2^{0.18M} = 123/12.9$ (power-law interpolation of Wetmur and Davidson[7], table 6 for the value at 0.18M).

2) The glucosylation of T4 (down triangles) DNA (Christiansen *et al.*[49], table 1) leads to multiplying the data by a factor $k_2^{100\%}/k_2^{0\%} = 1.85/1.23$.

3) DNA-RNA hybridization (squares) in comparison to DNA-DNA renaturation (Hutton and Wetmur[48], table 1) leads to multiplying the data by a factor $k_2^{DNA-DNA}/k_2^{DNA-RNA} = 1.9/1.3$.

4) Hydroxyapatite chromatography increases the measured rate in comparison (Miller and Wetmur[50]) with the UV absorbance technique and the corresponding data (diamonds) must be multiplied by a factor 0.6, in addition to the correction made above for the salt concentration of 0.18 M Na$^+$.

When these various corrections are applied, the data collapse to a unique curve, described by $k_2 = [(6.3 \pm 0.6) \times 10^5 \text{ mole}^{-1}\text{sec}^{-1}] \times N^\alpha$, with $\alpha = 0.51 \pm 0.01$.

The following comments can be made on these results:

1) The data collapse is observed with nucleic acids having different base compositions, with GC percentages ranging from 34% (for T4 and T2 DNAs) to 50% (for T7 and *Escherichia coli* DNAs). This agrees with the general conclusion that the base composition has an negligible effect on the reaction rate[1,51].

2) As stated above, the scaling law is observed with DNA (and RNA) fragments of different sizes obtained by a random degradation process, using either prolonged exposure to high temperature, shear in capillary tubes or sonication. This procedure yields a polydisperse population of single-stranded chains, whose average degree of polymerization <N> is measured by alkaline sedimentation. As a result of this fragmentation procedure, for a population with a given average <N>, the rate measured is a mean value obtained for an ensemble of molecules coming from varying regions of the initial genome and, therefore, is not sequence dependent. Note that, although the length of the reacting chains may be broadly distributed, the measured $k_2$ is strongly peaked at its average value. Indeed, since the measured rate is the sum average obtained from all reacting single-strands, because of the central limit theorem, its variance goes to zero as $M^{-1/2}$ where $M$ is the macroscopic number of chains in the solution. The measured rate for such random fragments is slower (by a theoretical factor 2/3) than the rate expected for the annealing of strictly complementary single-strands[7,52] (such as those produced with restriction enzymes).

3) The scaling law is valid not only for DNA-DNA hybridization, but also DNA-RNA hybridization (using bacteriophage $\varphi$X174 sequences[48]). The scaling law therefore does not depend on the chemical nature of the backbones of the reacting chains (ribose or deoxyribose sugars).

4) The scaling law is observed to hold true not only at 1 M NaCl[7] but also at 0.4 M NaCl[48,53], 0.18 M $Na^+$ (0.12 M Phosphate buffer)[33] as well as in 2.4 M TEACl (tetraethylammonium chloride)[47].

5) Finally, the scaling law is found to remain valid not only within the region around 20°C below the melting temperature where the apparent rate is maximal, but, in fact, over a wide range of temperature between 5 – 30°C below $T_m$[53].

*6) Conformation of the reacting single-strands: Studier's contribution revisited*

The conformation of the reacting polymers is clearly a key parameter of the reaction. Wetmur and Davidson made the assumption that the reacting single-strands have a strictly Gaussian conformation, their characteristic size $R$ thus being related to their degree of polymerization $N$ by the scaling law $R \propto N^{1/2}$. However, this assumption lacks experimental support and an analysis of the available data allows us to reach a different conclusion. This problem of the conformation of the reacting single-strands in thermal DNA renaturation was indeed investigated by Studier[8,54]. According to his work two types of intramolecular interactions within single-stranded chains can be distinguished: stacking interactions, which occur in the absence of (G-C and A-T) hydrogen bonding, and folding interactions, which involve hydrogen bonding (as well as stacking). Folding is due to hairpin structures, which are sequence-specific, and the conformation of a single-stranded polynucleotide is therefore expected to depend in a complex manner on its sequence. Eigner and Doty[55] independently proposed a terminology reflecting this idea, coarsely distinguishing between single-strands above $T_m$ and denatured strands below $T_m$. It is possible to refine this distinction if one takes into account the stability of all the possible hairpins that can exist within a single-stranded polynucleotide. Let $T_m^{hairpin}$ denote the highest melting temperature of all these hairpins. If $T_m^{hairpin} = T_m$ (the melting temperature of the corresponding DNA duplex) then, clearly, the dichotomy introduced by Eigner and Doty is valid and there is no temperature below $T_m$ where one can find the single-stranded polymer completely unfolded. This will be the case if the sequence contains a long palindrome, for instance for the synthetic polynucleotide polyd(A-T) studied experimentally by Inman and Baldwin[56] and theoretically by de Gennes[57]. If, however, $T_m^{hairpin}$ is strictly smaller than $T_m$, there exists a temperature range $T_m^{hairpin} < T < T_m$ where each complementary single-strand is in an unfolded conformation and the two strands can still renature.

The careful experimental work of Studier for bacteriophage T7 DNA shows that this second situation prevails under conditions of thermal renaturation:

- Studier observes that folding is extremely sensitive to changes in ionic strength and temperature which have, as well, a major influence on the renaturation of DNA. The fact that the scaling law given by equation 1 is valid over a broad range of salt concentrations and temperatures immediately shows that the details of folding cannot play a significant role in the scaling exponent.
- Combining absorbance and sedimentation studies of single-stranded DNA with renaturation kinetics in an elegant and astute manner, Studier shows that, when the temperature is increased at a fixed NaCl concentration, the reacting single-strands become completely unfolded at the point where the apparent maximal rate of renaturation is reached (that is, about 30°C below the melting temperature), and remain unfolded at higher temperatures.

These key observations allow us to conclude that, in the case of thermal renaturation of bacteriophage T7 DNA, at least, the reacting single-strands are free of hairpins and that their overall conformation can be described independently of their sequence in terms of universal good solvent conditions. In other words, we predict that the relationship between the degree of polymerization of such single-strands and their characteristic size, $R$, measured by an appropriate technique, will be described by the scaling law $R \propto N^\nu$, $\nu$ being the swelling exponent of good solvent conditions ($\nu \approx 0.588$), not that of $\Theta$ conditions ($\nu = 0.5$). Note that folding can also be eliminated under alkaline conditions, and Studier has shown that at high pH, the length dependence of the sedimentation coefficient $S$ is given by[58]: $S \propto N^{1-\nu} \propto N^{0.4}$ which corresponds to a swelling exponent of 0.6 as expected for good solvent conditions.

In summary, for all the cases where the scaling law given by equation 1 is observed, the reacting single-strands, whether ribo- or deoxyribonucleotides, are predicted to be in universal good solvent conditions without any significant folding. This general conclusion agrees with the finding that, under thermal conditions, the conformation of single-stranded homopolynucleotides is essentially determined by the rigidity of backbone, not by the bases. Furthermore, the rigidity of the backbone is essentially the same in ribo- and deoxyribonucleotides[59]. Similar conclusions concerning the lack of folding of reacting single-strands at high temperatures have also been reached for specific oligonucleotides[60]. Note finally, for the homopolynucleotide poly(riboadenylic acid), $\Theta$ conditions have been characterized by Eisenberg and Felsenfeld[61]. The locus of $\Theta$ points in the temperature–salt concentration plane of coordinates lies on a parabolic curve and not within a broad region of the phase diagram as required by the experimental observations considered here.

**Theory**

Let us now examine the theoretical models that can account for the observed results. In the absence of an energy barrier, the renaturation or hybridization reaction must be understood as a diffusion-controlled encounter reaction. In contrast, the presence of an energy barrier is compatible with two models, the first provided by the Transition State Theory and the second by a Kramers' process. As the Transition State Theory assumes thermodynamic equilibrium everywhere in space, and, in particular, no transport in the reaction, it therefore cannot account for the viscosity dependence of the reaction. In the following section, we shall see that the length dependence of the nucleation step can help to discriminate between the different possible processes.

*1. Equilibrium statistical mechanics of the interpenetration of the single-stranded strands*

The renaturation reaction proceeds in two steps: nucleation and zipping. For unique sequences, there are $N$ nucleation sites per chain of $N$ monomers, and the relation between $k_2$ and $k_{nu}$ is given by equation 4, namely $k_2 \propto k_{nu}N$. Our task is to understand the effect of strand length on the nucleation event. Nucleation requires the interpenetration of two complementary chains, here to be considered under dilute, good solvent conditions. The height of the barrier to be overcome during the reaction includes the energetic cost of the overlap, and the problem is to understand its length dependence.

*1.1 Interpenetration of polymer coils in good solvent*

We consider, therefore, the interpenetration of two polymer coils in dilute good solvent conditions. Each chain is to be made of $N$ identical monomers, with a radius of gyration $R_g \propto N^\nu$ and occupies a volume $V \propto R_g^d$, $d$ being the space dimensionality, thus $V \propto N^{\nu d}$. The energy of interaction $E_{int}$ is proportional to the average number of contacts between monomers of the two chains in the region of overlap (assumed in a situation of strong overlap to be of order $V$, the volume occupied by one chain).

In a mean-field approach[11,62], the polymers are viewed as clouds of uncorrelated monomers evenly distributed in the volume $V$. The energy $E_{int}$ is then proportional to the thermal energy $k_BT$ times the square of the average monomer concentration in the coils $n \approx N/V$, times the volume of overlap $V$

$$E_{int} \propto k_B T n^2 V \propto k_B T N^{2-\nu d} \tag{8}$$

This relation shows that, in the mean-field approach, very large chains are impenetrable (since with $d = 3$ and $\nu \approx 0.588$, $E_{int}$ scales as $N^{0.24}$). The equilibrium probability of overlap, given by the Boltzmann factor $e^{-E_{int}/k_B T}$ is expected to decrease exponentially with $N^{0.24}$. This approach is, however, invalid: by neglecting the correlation between the monomers, it greatly overestimates $E_{int}$. In reality, the energy of strong overlap is of order $k_B T \times \log N$ as explained below. This weak dependence implies that situations of strong overlap, such as those that can arise in the nucleation step, are possible, even for very large chains.

*1.2 Contact exponents in chemically-controlled reactions*

To obtain the correct length dependence for the nucleation step, it is necessary to take into account the contact exponents introduced by des Cloizeaux[63-65]. To do so we follow the approach of Khokhlov[66] in his analysis of chemically-controlled polymer-polymer reactions, i.e. slow reactions that can be described within the framework of the Transition State Theory[19]. Once again, the chains are assumed to be in good solvent conditions. Three types of reactions are considered: In the first type both reacting groups are located at the ends of the reacting chains; in the second type one group is at the end of the first chain and the second group in the central part of the second chain; in the third type both groups are located in central parts of the chains (see figure 3). The length dependence $k(N_1, N_2)$ of these reactions can be derived from the knowledge of the partition functions of the reacting chains (with degrees of polymerization $N_1$ and $N_2$ in the general case) and of the complex made by the two chains in contact. Each type of reaction is associated with a specific contact exponent[63] $\theta_i$. For the last type of reaction, in the case where $N_1 = N_2 = N$, the predicted length dependence is given by:

$$k(N) \propto N^{-\nu \theta_2} \qquad (9)$$

where the exponent $\theta_2$ is equal to[67] $0.82 \pm 0.01$. The power law predicted by equation 9 corresponds to an entropic effect and results from a logarithmic dependence of the overlap free energy.

*1.3 Application to hybridization and renaturation reactions*

In the case of DNA renaturation and hybridization, the first type of reaction corresponds to a nucleation step taking place between complementary sequences located at the ends of the two chains (figure 3a). The second type (figure 3b) corresponds to a nucleation step between the end of one chain and a complementary sequence located in the central part of the second chain (this type of nucleation event can exist because of the random degradation scheme employed). The third type corresponds to a nucleation step between two complementary sequences remote from the chain ends (figure 3c). Equation 9 assumes that there are no correlations with end effects (which are associated with the des Cloizeaux exponents $\theta_0$ for end-to-end contacts and $\theta_1$ for end-to-central contacts). In the case of hybridization and renaturation reactions, since the single-strands are both long and highly flexible with a persistence length of only a few monomers, central-to-central monomer contacts constitute the vast majority of the nucleation events. Equation 9 can thus be applied to the rate limiting nucleation step assuming that the Transition State Theory is valid:

$$k_{nu} \propto N^{-\nu\vartheta_2} \qquad (10)$$

which, in turn, yields:

$$k_2 \propto N^{1-\nu\vartheta_2} \propto N^{0.52} \qquad (11)$$

Equation 11 predicts correctly the measured length-dependence, as noted previously[68]. However, as mentioned above, the observed viscosity dependence of the reaction is incompatible with the assumptions of the Transition State Theory but remains compatible with either a diffusion-controlled or with a Kramers' process[20].

*2. Diffusion-controlled processes*

*2.1 Aggregation of colloids*

We examine now the situation of a diffusion-controlled encounter process between the two complementary single-strands. The concept of a diffusion-controlled reaction was introduced by Smoluchowski[12] to describe the aggregation of similarly-charged lyophobic, spherical colloids in the presence of added salt. In this case, the addition of salt screens the electrostatic repulsion and also destabilizes the colloids. When sufficient salt is added to the solution, the rate of aggregation becomes constant and is controlled by diffusion. The Smoluchowski equation for the steady-state value of the bimolecular rate constant $k_2^{aggr}$ is then given by:

$$k_2^{aggr} = 16\pi D_{coll} R_{coll} \tag{12}$$

where $D_{coll}$ is the diffusion coefficient of the colloid, and $R_{coll}$ its radius. Using the Sutherland-Einstein equation, $D_{coll} = \dfrac{k_B T}{6\pi \eta R_{coll}}$, one obtains:

$$k_2^{aggr} = \frac{8 k_B T}{3\eta} \tag{13}$$

This equation predicts that the rate of aggregation does not depend on the size of the colloids, and, indeed, this is observed experimentally. The other predictions are: 1. The independence of the rate on the concentration of salt; 2. A weak dependence on the temperature (through the temperature itself in the numerator and that of the viscosity in the denominator); and 3. A viscosity dependence, $k_2^{aggr} \propto \eta^{-1}$. For diffusion-controlled reactions involving small molecules, equation 12 can be used replacing the radius of the colloid $R_{coll}$ by a radius of capture $b$. The physical situation usually differs from that considered explicitely by Smoluchowski since, in general, there is no ongoing aggregation. Furthermore, the radius of capture $b$ can be much smaller than the overall size of the reacting molecules and the measured rates can therefore be much smaller than those predicted by equation 13 ($6 \times 10^9$ M$^{-1}$s$^{-1}$ in water at room temperature) and still be compatible with a diffusion-controlled process[13,14]. This is, in fact , exactly the situation here. Renaturation rates shown in figure 3 are in the range of $10^6$ - $10^8$ M$^{-1}$s$^{-1}$, a factor of 10 to $10^3$ slower than expected by equation 13.

*2.2 Diffusion-controlled processes involving polymers*

In renaturation and hybridization reactions, the reacting groups belong to long polymers. The analysis of diffusion-controlled processes involving polymers is delicate. Wetmur and Davidson[7] hypothesized that the radius of capture $b$ does not depend on the degree of polymerization $N$. This reasoning leads to the following length dependence for the nucleation rate in the diffusion-controlled case:

$$k_{nu} = 16\pi D_{SS} b \propto \frac{k_B T N^{-\nu}}{\eta} \tag{14}$$

where $D_{SS}$ is the diffusion coefficient of the single-stranded chain, which in turn yields the bimolecular rate constant:

$$k_2 \propto k_{nu} N \propto \frac{k_B T N^{1-\nu}}{\eta} \tag{15}$$

This reasoning leads to the result, already stated, that the observed length dependence would be compatible with a diffusion-controlled encounter reaction involving chains in a Θ solvent ($\nu = 1/2$), but not in a good solvent ($\nu = 0.588$), given the accuracy of the measurement of the scaling exponent $\alpha$.

The flaw in the previous hypothesis is that for polymeric chains, the reaction rate depends in a crucial manner on the short time-scale behavior of the monomer dynamics, as shown by Wilemski and Fixman[69,70], Doi[71,72], de Gennes[73], Grosberg *et al* [74] and others. When two chains in good solvent collide, they interpenetrate with an overlap volume of the order of the volume of the chains themselves (as shown by the length dependence of the energy $E_{int}$, discussed above). The overlap time $\tau_{ov}$ is proportional to $R^2/D_{ss}$, while the probability $p$ of a reaction per unit time is given by[71]:

$$p \propto \frac{1}{\tau_{relax}} \propto \frac{D_{SS}}{R^2} \tag{16}$$

where $\tau_{relax}$ is the longest relaxation time of the coil, here the Zimm relaxation time[75], $\tau_{Zimm}$. The reaction rate, here the nucleation rate, is a time-independent second-order rate constant, independent of $N$:

$$k_{nu} \propto p\tau_{ov} \propto \frac{\tau_{ov}}{\tau_{relax}} \propto 1 \qquad (17)$$

Equation 17 provides the asymptotic form of the nucleation rate constant, in the limit of large degree of polymerization, to be discussed below. Accordingly, the radius of capture $b$ would scale with the size of the coil and not be independent of $N$, as suggested by Wetmur and Davidson[7]. In fact, one recovers the scaling observed for the problem of colloidal aggregation. Equation 17, in turn, predicts that $k_2$ would scale linearly with $N$, in contradiction with the experimental results.

Equation 17 can be understood using the terms of compact and non-compact exploration introduced by de Gennes[73,76]. The problem considered by de Gennes is that of a diffusion-controlled encounter reaction between two reacting groups attached to polymers. The rate constant is shown to depend on the root mean squared (rms) displacement $x(t)$ of one reacting group during a time $t$. The situation where the quantity $t^{-1}x^3(t)$ increases with time corresponds to a regime of non-compact exploration; it is obtained in the classical case where the reacting groups belong to small molecules where simple diffusion prevails ($x(t) \propto t^{1/2}$). This leads to a well-defined, time-independent second-order rate constant. The opposite situation, where $t^{-1}x^3(t)$ is strictly decreasing with time, corresponds to a regime of compact exploration. The rate constant in that case, proportional to $t^{-1}x^3(t)$, decreases with time. The cases examined by de Gennes were that of dense polymer systems, both non-entangled[73] and entangled[76]. This description can also be applied to the situation at hand, that of dilute non-entangled polymers[77]. For single-stranded DNA chains, the rms displacement of the monomers in dilute solutions has been investigated recently[78]: it is proportional to $t^{1/3}$ up to the longest relaxation time ($\tau_{Zimm}$) of the chain (as predicted for single chains with hydrodynamic interactions by Dubois-Violette and de Gennes[79]). The quantity $t^{-1}x^3(t)$ is therefore non decreasing (constant). In conclusion, a diffusion-controlled encounter reaction between groups attached to single-stranded DNA chains should correspond to a situation of non-compact exploration on short time scales, leading to the time and length-independent reaction rate given by equation 17.

*Kramers' theory*

Since a diffusion-controlled encounter process is to be ruled out, we finally examine Kramers' approach. Kramers' theory describes chemical reactions in terms of the barrier crossing by a Brownian particle. Although it was originally derived using a one-dimensional Langevin equation, it can be generalized to higher dimensional reaction coordinates, provided the relevant barrier is a saddle-point with a unique unstable direction.

Kramers' theory describes the diffusion of a particle in a double-well potential, in the limit where the barrier height is much larger than the thermal energy $\Delta E_b \gg k_B T$. This hierarchy of energies implies a hierarchy in time scales without which a rate description is meaningless. Indeed, in the case when $\Delta E_b \approx k_B T$, the reaction is diffusion-controlled and Kramers' theory does not apply.

Kramers' theory starts from the Langevin equation, that is Newton's second law with a friction force and a random noise

$$m\frac{d^2x}{dt^2} + \gamma\frac{dx}{dt} + \frac{\partial U}{\partial x} = \xi(t) \qquad (18)$$

where $\gamma = k_B T/D$ is the friction coefficient, proportional to the solvent viscosity $\eta$ ($D$ is the diffusion coefficient), and $\zeta(t)$ is a white Gaussian noise with zero average and having a correlation function given by

$$\langle \zeta(t)\zeta(t')\rangle = 2\frac{k_B T}{D}\delta(t-t') \qquad (19)$$

The above relation ensures that in the long time limit, the Langevin dynamics samples the Boltzmann distribution (ergodicity). Assuming high barriers and harmonic-like potentials around the reactants, products and transition states, Kramers has computed the reaction rates in various regimes that we discuss now.

- In the medium to high viscosity regime $\gamma > \omega_b$, the Kramers' rate formula is:

$$k = \frac{\omega_0}{2\pi\omega_b}\left[\sqrt{\frac{\gamma^2}{4}+\omega_b^2} - \frac{\gamma}{2}\right]e^{-\frac{\Delta E_b}{k_B T}} \qquad (20)$$

where the subscript 0 refers to the initial or final state of the system, $\omega_0$ denotes the normal mode frequency in the well, $\omega_b$ that of the unstable mode at the top of the barrier, and $\Delta E_b$ that of the energy barrier between the state 0 and the top of the barrier.

- In the high viscosity regime $\gamma \gg \omega_b$, the above rate can be expanded in $\omega_b/\gamma$, yielding one of the celebrated Kramers' formulas:

$$k = \frac{\omega_0 \omega_b}{2\pi\gamma}e^{-\frac{\Delta E_b}{k_B T}} \qquad (21)$$

Note that in this high viscosity regime, the reaction rate is inversely proportional to the solvent viscosity $\eta$.

- In the low viscosity regime $\gamma \ll \omega_b$, the Kramers' formula (20) is no longer valid since the energy of the system is almost conserved. The reaction rate is given by another formula (also due to Kramers):

$$k = \frac{\omega_0}{2\pi} \frac{\gamma}{k_B T} I(E_b) e^{-\frac{\Delta E_b}{k_B T}} \qquad (22)$$

where $I(E_b)$ is the value of the action $I = \oint p\, dq$ at the barrier energy. Note that in the low viscosity regime, the reaction rate is proportional to the friction coefficient in sharp contrast with the high viscosity case.

In the long time limit, one expects the inertial term to be negligible compared to the friction term. The time scale $\tau$ beyond which this occurs can be obtained by a simple scaling argument: Equating the first two terms in (18), which scale respectively as $m/\tau^2$ and $\gamma/\tau$, one finds $\tau = m/\gamma$. Thus, for times $t \gg \tau$, the inertial term in (18) can be neglected, and the system is in the high viscosity regime.

For DNA, the typical mass of a nucleotide is $m \sim 10^{-25}$ kg and the estimate for its diffusion constant is $D \sim 10^{-11}$ m$^2$/s so that beyond a time scale of the order of $10^{-15}$ s, the inertial term in equation 18 is negligible, and we are in the high viscosity regime where the reaction rate is given by equation 21.

For the specific case of renaturation or hybridization, equation 21 applies to $k_{nu}$. For this type of complex molecular reactions, $\Delta E_b$ should be replaced by the free energy barrier $\Delta F_b = \Delta U_b - T\Delta S_b$. According to equation 9, the length dependence of the reaction is solely due to the entropic contribution $\Delta S_b$.

In Kramers' theory as in the transition state theory (TST), the rates have a Boltzmann exponential dependence on the barrier energy. The major difference between the two theories is in the dependence of the rates on the solvent viscosity $\eta$. In TST, the system is assumed to be at equilibrium everywhere in space, and the friction plays no role, whereas in Kramers' theory, there is a stationary flux over the barrier, the magnitude of which is proportional to the inverse of the friction coefficient. Another major assumption of the TST is that once a system crosses the transition state, it goes irreversibly to the product state, whereas in Kramers theory, the system undergoes Brownian motion, and the probability for the particle to go to the product or reactant state is equal to ½ at the transition state. As a result, TST overestimates transition rates, and this becomes particularly important at large friction. Note that the friction coefficient entering Kramers' formula has to be understood as the total friction coefficient, namely the sum of the friction of the molecules with the solvent as well as an internal friction. This internal friction may be an important slowing factor for dense polymeric systems, particularly in the dense globular, phase of proteins where the solvent is expelled from the core of the system[25]. However, this is not the case here for renaturation and hybridization reactions since the reacting single-strands are in a good solvent and the total friction reduces to that of the solvent.

In the context of a Kramers' process the nucleation rate $k_{nu}$ can be written:

$$k_{nu} = \frac{B \times N^{-\nu \theta_2}}{\eta} \qquad (23)$$

where $B$ is a constant independent of the viscosity and of the length of the reacting strands, and where the length dependence obeys the prediction made by equilibrium statistical mechanics (equation 9). The power law is associated with the entropic barrier described above. The viscosity dependence of the reaction also implies that the activation energy of the solvent contributes to the temperature dependence of the reaction rate. From the value of 30 kJ.mol$^{-1}$ determined by Thrower and Peacocke[5,6], one must subtract 13 kJ.mol$^{-1}$ corresponding to the temperature dependence of the water viscosity in this temperature range. The final activation energy $\Delta E_b$ is thus equal to 17 kJ.mol$^{-1}$, about 7 $k_B T$. Such an activation energy is sufficiently high to probe the equilibrium properties of the reacting polymers.

We conclude that a Kramers' process is the only model compatible with the data. The reaction combines both Brownian motion and barrier crossing, and the length dependence of the reaction can be determined from equilibrium statistical mechanics.

*Range of validity of the Kramers' process*

Equation 23 provides an adequate description of the experimental results for intermediate time scales (between 10 to 10,000 seconds) and for the experimentally accessible range of the degree of polymerization (100 ≤ $<N>$ ≤ 50,000). Is it still valid asymptotically for long times and for long chains? This is a delicate question:

1) We discussed above the slowing down of the reaction rate at long times due to the existence of single-stranded dangling ends. Experimentally, this behavior could be bypassed to a limited extent by using denatured double-stranded restriction fragments of various lengths.

2) DNA renaturation is an irreversible bimolecular reaction in which exactly equal quantities of complementary chains combine to form an inert double helix. Formally, this corresponds to a situation of trapping or of particle-antiparticle annihilation[80-82]. The time asymptotics of such bimolecular diffusion-controlled reactions are expected to differ from that predicted by equation 3: density fluctuations lead to a spatial segregation of the reactants and to a slowing down of the reaction. This implies that equations 12 and 17 describing the rate constants cease to be valid for very long time. Because fluctuations are also central to a Kramers' process, the same limit of validity is expected to apply as well to equation 23.

3) Friedman and O'Shaugnessy[77] have studied theoretically the asymptotic behavior of an intermolecular polymer reaction in dilute, good-solvent conditions. They concluded that the diffusion-control limit does not, in fact exist; for very long chains the rate of the reaction should then scale as predicted by equation 11. However it is not clear whether this asymptotic behavior is compatible with the viscosity dependence predicted by equation 23. Furthermore, the issue of polymeric entanglement, which could be a crucial problem for very long chains, has also to be considered.

We can therefore conclude that equation 23 has a limited range of validity and it will not apply to very long times and to very long chains.

**Discussion**

*A Kramers' process in Watson-Crick base pairing*

The purpose of this work has been to try to clarify of the mechanism of renaturation and hybridization reactions under thermal conditions. Our conclusion is that they can be solely described by a Kramers' process. Transition State Theory does not account for the viscosity dependence; a diffusion-controlled process does not account for the length dependence. It is important in this respect to underline that the term "diffusion-controlled" is often used in the biophysical literature in a very broad manner and most often includes Kramers' processes which are only diffusion-assisted. The present work shows that it is indeed possible to discriminate between these two processes. Another example closely related to the reactions studied here is that of the random coil to helix transition in the homopolynucleotide poly(riboadenylic acid)[83]. The rate of helix formation is inversely proportional to the solvent viscosity, and its activation energy is also about 17 kJ.mol$^{-1}$, or about 7 $k_B T$. The reaction is described by Dewey and Turner[83] as "a rotational diffusion controlled reaction", but is probably also a Kramers' process.

*Biological specificity and physical universality in Watson-Crick base pairing*

Specificity through complementary recognition is a hallmark of molecular biology. Our understanding of biological specificity is based on a detailed, local knowledge of the structure of the interacting biomolecules. Watson-Crick base pairing in nucleic acids is an instance of this highly specific interaction. This route to understanding contrasts with the approach to physical phenomena based on universal concepts. In physics, the explanation of universality relies on global symmetry consideration and leads to dimensional analysis, scaling and fractals[84-86]. Scaling concepts have been introduced with great success in the study of critical phenomena[87] and in polymer physics[38]. The concepts of biological specificity and physical universality are seemingly antinomical, and it is therefore of interest to elucidate the circumstances where they apply simultaneously, as is the case in hybridization and renaturation reactions of polymerized nucleic acids. Indeed, under thermal conditions they have an extreme specificity, while at the same time their length dependence is described by a universal scaling law, independent of the specific sequences involved.

The present work indicates that the observation of a universal scaling law in renaturation and hybridization reactions is made possible by two facts. First, the reacting complementary single-strands are both in a universal, good solvent regime, yet this remains compatible with the complementary recognition. The scaling law has its origin in the universality of the solvent conditions. Second, the random degradation procedures employed in the experiments prevents the possible blurring of the scaling law by sequence effects that might occur with the use of restriction fragments. Specificity (through Watson-Crick base pairing) and universality (through good solvent conditions and random degradation) are not antinomical in this situation. Rather, both concepts have to be taken into consideration simultaneously to understand the biological process.

*Related problems*

In the present work we have investigated the mechanism of renaturation and hybridization reaction under simple (thermal) conditions. It is hoped that the ideas developed here will find applications to the understanding of more complex conditions (addition of inert polymers, use of heterogeneous systems, immobilization of one of the reacting single-strand, excess of one of the complementary single-strand over the other). We mention here three illustrations.

1. Catalysis of these reactions by various chaperones: In the presence of *Escherichia coli* Single Strand Binding Protein, the scaling law given by equation 1 is also observed[88], suggesting that here as well the complementary single-strands are in an fully unfolded state, but at a much lower temperature (37°C). This could be tested experimentally.

2. It has been shown that the coupling of DNA renaturation with an aggregation process can greatly speed up thermal renaturation[89]. Here we find a physical situation similar to the Smoluchowski problem of colloidal aggregation: there is an ongoing aggregation process and the chains renature (or hybridize) at the same time. Under these heterogeneous conditions, the values of the measured initial bimolecular reaction rates are closer to the prediction of equation 13 for a diffusion-controlled reaction. A variety of ssDNA aggregating agents have been shown to increase the reaction rates, including monovalent salts at salting-out concentrations[90], multivalent cations[89,91,92], or cationic surfactants[93]. The ideas developed here could serve to clarify the mechanism of these more complex reactions.

3. Hairpin formation is intensely studied. Whether these reactions are diffusion-controlled or not is unknown[94], and could be clarified through a study of their length dependence.

*The work of Wetmur and coworkers*

The work of Wetmur and his coworkers has been highly influential, and it is appropriate to summarize its strong points and shortcomings. The experimental validity of the beautiful scaling law discovered by Wetmur and Davidson is undisputable. They correctly interpreted their result as an excluded volume effect, according to which the increasing size of the single-stranded chains decreased the accessibility to nucleation sites. However:

1) The analytical estimate of this effect is incorrect. Wetmur and Davidson (and before them Wada and Yamagami[95]) used a simple Flory-Krigbaum[62] mean field approach now known to be invalid. The same criticism applies to ulterior work of Wetmur on this subject[96]. Furthermore, the correct scaling in this approach is given by combining equations (4) and (8)

$$k_2 \propto N e^{-E_{int}/k_B T} \propto N e^{-N^{2-\nu d}} \qquad (24)$$

and not by

$$k_2 \propto N^{\nu d - 1} \qquad (25)$$

as predicted by Wetmur and Davidson[7] and by Sikorav and Church[89]. Equation 25 arises from a confusion between the energy $E_{int}$ given by equation 8 and the corresponding Boltzmann factor. As stated above, the length dependence of $k_2$ given by equation 10 implies a logarithmic dependence for the overlap energy, which cannot be obtained with a mean-field theory.

2) The assumption that the single-stranded chains are at the $\Theta$ point[7] is incompatible with the experimental data[54].

3) Wetmur and Davidson have suggested that the viscosity dependence of the reaction was due to the zipping step. This proposal is unlikely, since the hydrodynamic drag on the rotating single-strands is negligible in the reaction[97].

**Conclusions**

We have developed a simple theoretical analysis of the mechanism of thermal renaturation and hybridization reaction that accounts for the available experimental data. This analysis leads to testable predictions, for instance that the reacting single-strands are in good solvent conditions. The scaling law described by equation 1 can also be checked with restriction fragments; this could require the presence of compounds suppressing the GC dependence of the helix-coil transition[98].

ACKNOWLEDGMENT We thank Arach Goldar for suggestions and comments on the manuscript as well as Gérard Jannink (who sadly passed away in August 2008) and Bertrand Duplantier for discussions. The comments of an anonymous reviewer have also contributed to improve the manuscript.

**Figure legends**

**Figure 1**. Helix-coil transition, renaturation and hybridization

Starting from a native linear DNA double helix (right), one can under appropriate conditions generate the two complementary single-strands (left). The transition between the two conformations is called the helix-coil transition. The disruption of the helical order is called denaturation. Renaturation is the reaction by which the two separated complementary single-stranded DNA chains anneal to generate the native DNA double helix. Hybridization reactions involve single-stranded nucleic acids that are not perfectly complementary, having for instances mismatches, different lengths, or different chemical natures (RNA or PNA (for Peptide Nucleic Acids) rather than DNA; giving rise for instance to DNA-RNA, RNA-RNA hybrids).

The simulation illustrates a DNA double helix of length ~ 150 nm ($N$ = 450), about 3 persistence lengths. The denatured DNA single strands are shown in configurations of random walks with excluded volume.

**Figure 2.** Bimolecular rate constant $k_2$ as a function of $<N>$

**a.** Up triangles: bacteriophage T7 DNA (1M $Na^+$)[7,52]; Circles: T7 (0.4 M $Na^+$)[48]; Down triangles: bacteriophage T4 DNA (1 M $Na^+$)[7,52]; Squares: bacteriophage $\varphi$X174 DNA-RNA hybrids (0.4 M $Na^+$)[48]; Diamonds: *Escherichia coli* DNA[33]. The mean ssDNA fragment length $<N>$ was calibrated from the mean sedimentation constant $<S>$ via the molecular weight according to Studier[58] as: $<M_w> = (<S>/0.0528)^{1/0.400}$ and $<N> = <M_w> \times N_0/(M_0/2)$, where $N_0$ and $M_0$ are the length and molecular weight of the unfragmented double-stranded DNA. The bimolecular rate constant is given here per unit of moles of chains, thus multiplying the data in the literature (given in moles of phosphate) by a factor of $2N_0$. The lines are the results of a nonlinear least-squares fit of a single power-law to the ensemble of the data: $(k_2)_i = A_i N^{0.502(9)}$ with $A_{T7} = 7.8(6) \times 10^5$, $A_{T4} = 4.3(3) \times 10^5$, $A_{T7,0.4M} = 3.5(3) \times 10^5$, $A_{\phi X174} = 2.4(2) \times 10^5$ and $A_{E.coli} = 1.11(8) \times 10^5 \, \text{mole}^{-1} \text{sec}^{-1}$.

**b.** Data rescaled to equivalent 1M Na$^+$ as described in the text. The data collapse to one curve: $k_2 = [6.3(6) \times 10^5 \, \text{mole}^{-1} \text{sec}^{-1}] \, N^{0.51(1)}$.

**Figure 3.** Nucleation in DNA renaturation

Three types of nucleation events are described: (a) Nucleation between complementary sequences located at the ends of two single-strands. (b) Nucleation between the end of one chain and a complementary sequence located in the central part of the second chain. (c) Nucleation between two complementary sequences remote from the chain ends. Shown are simulations of a random walk with excluded volume as in figure 1 as well as a schematic sketch of the chains where the nuclei are drawn as thicker lines.

**Figure 1**

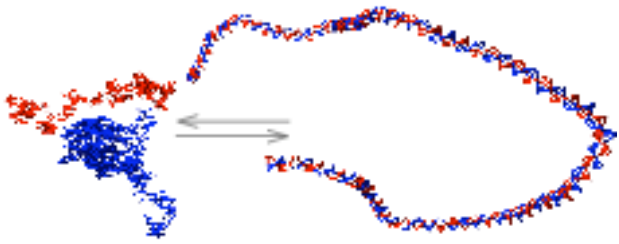

**Figure 2**

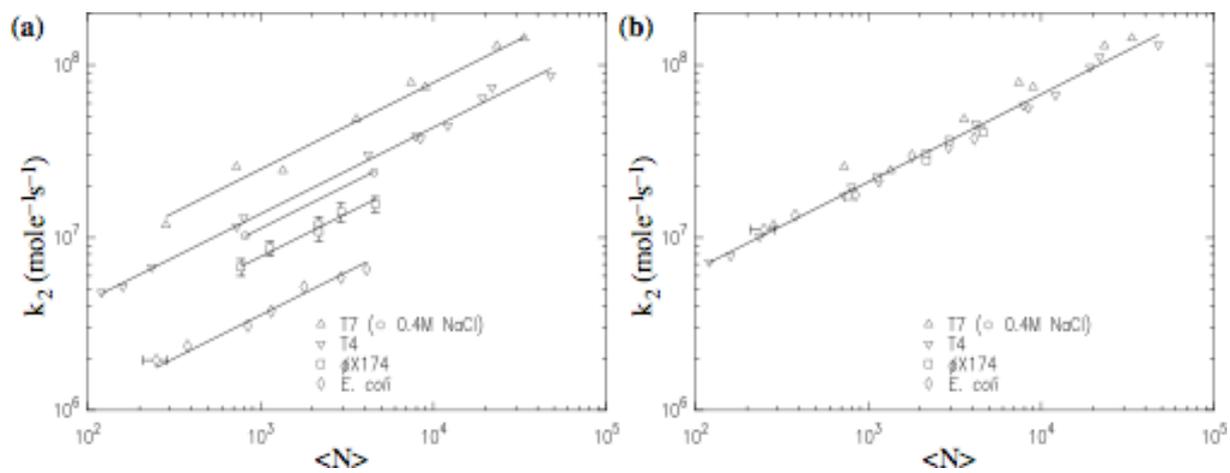

**Figure 3**
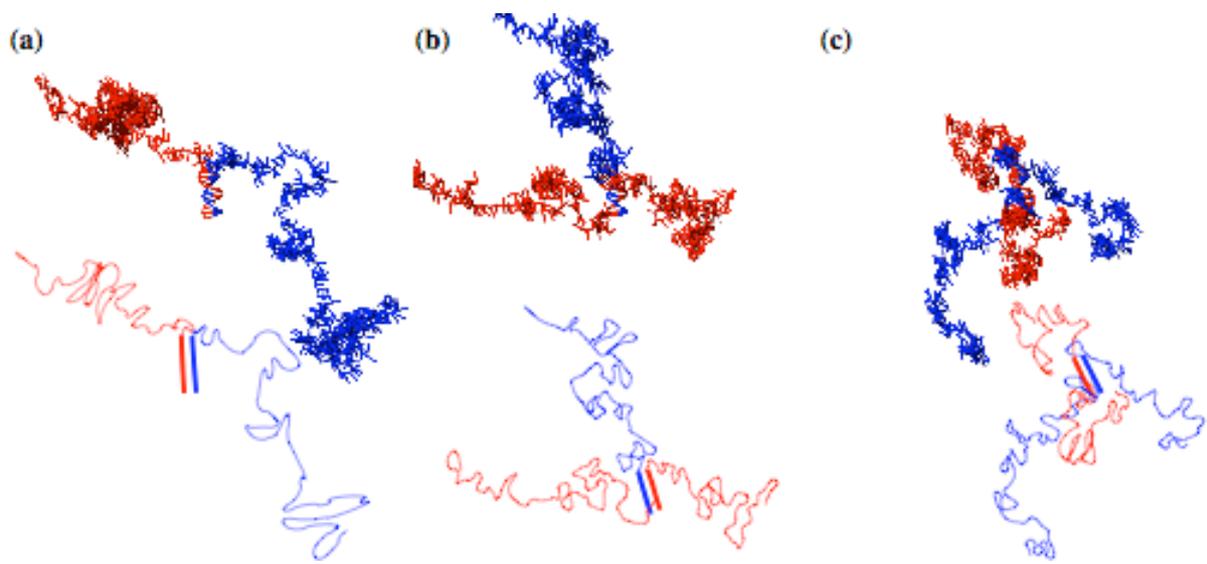